\documentclass[pre,groupedaddress,preprint]{revtex4}
\usepackage{amsmath}
\usepackage{amssymb}
\usepackage{color}
\usepackage[dvips]{graphicx}
\usepackage{tabularx}

\makeatletter

\newcommand{\Rmnum}[1]{\expandafter\@slowromancap\romannumeral #1@}
\makeatother
\newcommand{\myv}[1]{\v{#1}}
\renewcommand{\v}[1]{\textbf{\textit{#1}}}
\renewcommand{\d}[1]{\textsf{#1}}

\begin{document}
\title{Optimizing working parameters of the twin-range cutoff method
  in terms of accuracy and efficiency}

\author{Han Wang} \affiliation{LMAM and School of Mathematical
  Sciences, Peking University, Beijing} \email{han_wang@pku.edu.cn}
\author{Pingwen Zhang} \affiliation{LMAM and School of Mathematical
  Sciences, Peking University, Beijing}

\begin{abstract}
  We construct a priori error estimation for the force
  error of the twin-range cutoff method, which is widely used to treat
  the short-range non-bonded interactions in molecular
  simulations. Based on the error and cost estimation,
  we develop a work flow that can automatically
  determine the nearly most efficient twin-range cutoff parameters
  (i.e. the cutoff radii and the neighbor list updating frequency)
  prior to a simulation for a predetermined accuracy. Both the error
  estimate and the parameter tuning method are demonstrated to be
  effective by testing simulations of the standard Lennard-Jones 6-12
  fluid in gas, liquid as well as supercritical state. We recommend
  the tuned twin-range cutoff method that can save precious 
  user time and computational resources.
\end{abstract}

\maketitle

\section{Introduction}
\label{sec:intro}
Non-bonded interactions are encountered in nearly
  every molecular simulation, but their calculations are computationally expensive.
It is thus important to develop computational methods that boost both efficiency and accuracy at the same time.
Non-bonded interactions are
mainly formed by a pairwise interaction $u(r)$, where $r$ is the distance
between two interacting particles. There are two types of pairwise interactions,
namely the long-range interaction and the short-range interaction, depending on
the rate at which $u(r)$ decay with respect to the distance.


Most short-range interactions satisfy $\vert
u(r)\vert\leq C r^{-m}$, $m>3$, which guarantees an absolute convergence of the
energy. A naive idea to treat short-range interactions
is to explicitly calculate and sum all pairwise interactions. This
results in a computational cost scaling $\mathcal O(N^2)$ per time step, which
rapidly becomes inefficient as the number of particles $N$ grows large. A
better way is to introduce a cutoff radius, outside of which all pairwise
interactions are simply neglected. In combination with the cell list and the
neighbor list algorithms \cite{frenkel02b}, the total computational cost of the
short-range interactions can be reduced to an acceptable level of $\mathcal
O(N)$.

It is well known that various physical properties
show significant dependence on the cutoff radius
and the method to treat the discontinuity at the
cutoff, for example, the phase diagram of the Lennard-Jones fluid
\cite{smit1992phase, baidakov2000effect, ou2005molecular}, the density profile
of the liquid-vapor interface \cite{panagiotopoulos1994molecular,
mecke1997molecular}, the surface tension \cite{trokhymchuk1999computer,
baidakov2000effect, shen2007comparative} and the free energy calculation
\cite{ou2005molecular}. Ill-chosen cutoff can lead to undesirable artifacts, for example,
  the phase diagrams of the Lennard-Jones fluid have been demonstrated to be
  substantially different with different choices of cutoff radii \cite{smit1992phase}.
A straightforward way to eliminate the cutoff
effects is to use an extremely large cutoff radius such that all properties of
interest satisfactorily converge. This has motivated the adoption of  a
non-truncated potential (i.e. the cutoff radius is the same as half of the
simulation box) to study the critical point phenomena \cite{lotfi1992vapour,
caillol1998critical, potoff1998critical, perez2006critical}. However, such a
long cutoff radius drastically increases the computational cost, thereby prohibiting long time
and large size molecular simulations. An alternative way is to use a small cutoff
while applying long range correction (LRC) \cite{allen87a} to eliminate the
systematic error of the potential energy, the pressure and the free energy. The
idea is to integrate the thermodynamic properties from the cutoff radius to
infinity, assuming the radial distribution function is equal to 1. Some more
sophisticated methods have been developed in recent years to improve LRC for the
constant pressure simulation \cite{lague2004pressure}, the inhomogeneous systems
\cite{janecek2006long} and the free energy calculation of systems containing
macromolecules \cite{shirts2007accurate}.

A promising way to   
  quantitatively analyze the undesirable cutoff effects is to
  express these artifacts in terms of the difference between the
  cutoffed interaction and the exact interaction.
There are several error analysis works on the
long-range interaction algorithms \cite{kolafa1992cee, hummer95a,
  petersen1995accuracy, deserno1998mue2, wang2010optimizing}, while
the analysis on the short-range interaction is scarce.
In this paper, we develop the  error estimate of the short-range
force introduced by the widely used twin-range cutoff method
(which is reduce the commonly used single-range cutoff method by a special choice
of parameters) in homogeneous
systems. 
Furthermore, to automatically determine the most efficient
cutoff radii and neighbor list updating frequency, a work
flow optimizing these working parameters with respect to speed and accuracy is
proposed and tested.

\section{Theory}
\label{sec:theory}
\subsection{The twin range cutoff method}


The twin-range cutoff method introduces two cutoffs, denoted by $r_1$
and $r_2$, with $r_1\leq r_2$.  Assuming the neighbor list updating
frequency is $M$, then at every $M$ steps, the neighbor list is built
for all neighboring particles that fall in the short cutoff radius $r_1$ and the corresponding
interactions are calculated and applied every next $M-1$ steps.  The neighbors
fall in between the short cutoff and the long cutoff are assumed to move slowly,
so the interactions can be calculated less frequently.
At every $M$ steps,
the interactions with these particles are
calculated and stored. In the subsequent $M-1$ steps, they are applied
to the corresponding particles
without any change.  When $r_1 = r_2$, the twin-range cutoff
method reduces to the single-range cutoff method.

\subsection{The error estimate of the twin-range cutoff method}
\label{sec:error}
For simplicity, we will only study the error estimate of single
component systems, because it is not difficult to derive the error estimate for
multicomponent systems in the same way.
We consider a reference particle $i$ and
denote the sets of  neighbors fall in $r_1$, between $r_1$ and $r_2$, and
out of $r_2$ by $\Omega_1^i$, $\Omega_2^i$ and $\Omega_3^i$, 
respectively.  The position of any particle $j$ is denoted by
$\v r_j$ at the step when the neighbor list is built. In the following
several steps, the accumulated absolute displacement is denoted by $\v
d_j$.  Then the exact force $\v F_i^\ast$ on particles $i$ and the cutoffed force
$\v F_i$ by the twin-range cutoff method are
\begin{align}
  \v F_i^\ast & = \sum_{j\in\Omega_1^i}\v F(\v r_{ij} + \v d_{ij}) +
  \sum_{j\in\Omega_2^i}\v F(\v r_{ij} + \v d_{ij}) +
  \sum_{j\in\Omega_3^i}\v F(\v r_{ij} + \v d_{ij}),\\
  \v F_i & = \sum_{j\in\Omega_1^i}\v F(\v r_{ij} + \v d_{ij}) +
  \sum_{j\in\Omega_2^i}\v F(\v r_{ij}).
\end{align}
Therein the relative position is $\v r_{ij} = \v r_i - \v r_j$ and the relative
displacement is $\v d_{ij} = \v d_i - \v d_j$.

The difference between the exact force and the cutoffed force is
therefore
\begin{align}\nonumber
  \Delta\v F & = \v F_i^\ast - \v F_i\\\nonumber & =
  \sum_{j\in\Omega_2^i}[\,\v F(\v r_{ij} + \v d_{ij}) - \v F(\v
  r_{ij})\,] + \sum_{j\in\Omega_3^i}\v F(\v r_{ij} + \v d_{ij})
  \\\label{tmpeq:5} & \approx \sum_{j\in\Omega_2^i}\nabla\v F(\v
  r_{ij})\cdot\v d_{ij} + \sum_{j\in\Omega_3^i}[\v F(\v r_{ij}) +
  \nabla\v F(\v r_{ij})\cdot\v d_{ij}].
\end{align}
The last approximation holds by Taylor expansions, so it is important
to keep in mind that~\eqref{tmpeq:5} is only valid when $\vert\v
d_{ij}\vert$ is small. To estimate the error of the force, it is
crucial to provide a proper definition of the word ``error'' first. We
adopt the widely used root mean squared (RMS) force error
$\mathcal E (r_1, r_2, M) = \sqrt{\langle\vert\Delta \v
  F\vert^2\rangle}$,
where the $\langle\cdot\rangle$ is the average over all positions $\v
r_i, i=1,\cdots,N$ and all displacements $\v d_i, i=1,\cdots,N$.

To
calculate this error we need some assumptions:
\begin{enumerate}\itemsep -5pt
\item $\v r_i,\ i=1,\cdots,N$ are random variables with uniform
  distributions over the space.
\item $\v d_i,\ i=1,\cdots,N$ are random variables having a normal
  distribution with mean 0 and variance $d^2$, namely $\mathcal{N}(0,
  d^2)$.
\item $\v r_i$ and $\v r_j$ are independent, if $i\neq j$.
\item $\v d_i$ and $\v d_j$ are independent, if $i\neq j$.
\item $\v r_i$ and $\v d_j$ are independent, for any $i$ and $j$.
\end{enumerate}
The standard deviation $d$ of the random variable $\v d_i$ can be
approximately related to the neighbor list updating frequency $M$ by
\begin{align}\label{tmpeq:7}
  d \approx M\Delta t\sqrt{\frac{3k_BT}m},
\end{align}
where $k_B$ is the Boltzmann constant, and $m$ is the mass of the particles.
By the theorem of equipartition, ${3k_BT}/m$ is the mean squared
velocity. This relation is only
valid when $d$ is smaller than the mean free path. In
all the test simulations studied in this paper, the largest deviation
of $d$ from $M\Delta t\sqrt{3k_BT/m}$ is less than 10\%.  All the
above assumptions are ideal cases that facilitate the derivation of the error
estimate. However, in systems studied in real problems, these assumptions
can be violated. In section \ref{sec:result}, we will show when the
real force error deviates from our theoretical estimate and to what
extent the theoretical estimate is reliable.

Based on the aforementioned assumptions, we reach the error estimate:
\begin{align}\nonumber
  \mathcal E^2 (r_1, r_2, M)
  = &
  (4\pi\rho)^2
  d^2 \bigg\{
  \int_{r_1}^{\infty}\frac13r^2u''(r)\d dr + 
  \int_{r_1}^{\infty}\frac23ru'(r)\d dr 
  \bigg\}^2 +  \\\nonumber
  &
  8\pi\rho\,
  d^2 \bigg\{
  \int_{r_1}^\infty\frac13\,[ru''(r)]^2\d dr + 
  \int_{r_1}^\infty\frac23\,[u'(r)]^2\d dr  
  \bigg\} + \\\label{error}
  &
  4\pi\rho\int_{r_2}^\infty [ru'(r)]^2\d dr.
\end{align}
It is straightforward to develop the error estimate of the short-range
interaction that has the form $u(r) = 4\varepsilon(\sigma/r)^{m}$ and
$m>3$:
\begin{align}\nonumber
  \mathcal E^2 (r_1, r_2, M)
   = \;&
  (\frac{16}3m\pi\rho\varepsilon\sigma^m)^2d^2
  \bigg(\frac1{r_1^{2m-2}}\bigg) +\\\nonumber
  &
  2\frac{m^2+2m+3}{3(2m+1)}
  \pi\rho(8m\varepsilon\sigma^m)^2 d^2
  \bigg(\frac1{r_1^{2m+1}}\bigg) + \\\label{error.sr}
  &
  \frac{4}{2m-1}
  \pi\rho(4m\varepsilon\sigma^6)^2
  \bigg(\frac1{r_2^{2m-1}}\bigg).
\end{align}

\subsection{Tuning the working parameters for the twin-range cutoff method}
\label{sec:opti}

In general, the calculation of the cutoffed
short-range interaction with the neighbor list method is performed in two steps:
1, generate the
neighbor list at every $M$ steps. 2, calculate the interactions by using the
neighbor list in the subsequent
$M-1$ steps. We use the following formulas to estimate the
computational costs of the force calculation ($T_{F}$) and the neighbor list
generation ($T_{N}$):
\begin{align}\label{tf}
  T_{F} &= c_1 r_1^3 + d_1,\\\label{tn} T_{N} &= c_2 r_2^3 + d_2.
\end{align}
Where $c_1$, $d_1$, $c_2$ and $d_2$ are constants depending on the
system being studied, the software implementation, as well as the hardware
architecture. For a good estimate of these constants, we time a series
of short test runs with different cutoff radii and fit the
computational costs in the least square sense. The average
computational cost of the short-range interaction per step is
\begin{align}\label{tsr}
  T = T_F + \frac1MT_{N}.
\end{align}

Provided with both the error and the computational cost estimates for the
twin-range cutoff method, it is possible to design a routine that
determines the most efficient combination of parameters $r_1$, $r_2$ and
$M$.  Here, the phrase ``most efficient'' refers to a set of parameters
that reaches a given accuracy at minimal computational cost.  From
a mathematical point of view, this is a constrained optimization
problem that can be written in the following way:
\begin{align}
  \label{opti1.min}
  \min\quad &T(r_1, r_2, M),\\
  \label{opti1.c1}
  \textrm{\textbf{s.t.}}\quad
  &\mathcal E(r_1, r_2, M) = \mathcal E^\ast \quad \textrm{and} \\
  \label{opti1.c2}
  &d\,(M) \leq d_0.
\end{align}
Where $\mathcal E^\ast$ defines the required accuracy. Constraint
\eqref{opti1.c2} is added because the Taylor expansions in
\eqref{tmpeq:5} are good approximations only when $d$ is small.  In the
multicomponent systems, the displacement should be constrained for the
lightest particles. We find $d_0 = 0.2\sigma$ will give reasonable
results. Since $M$ can be analytically solved by constraint
\eqref{opti1.c1}, saying $M = M(r_1, r_2, \mathcal E^\ast)$,
\eqref{opti1.min} -- \eqref{opti1.c2} is reduced to
\begin{align}
  \label{opti2.min}
  \min\quad & T(r_1, r_2, M(r_1, r_2, \mathcal E^\ast)),\\
  \label{opti2.c}
  \textrm{\textbf{s.t.}}\quad &
  d\,(M(r_1, r_2,\mathcal E^\ast)) \leq d_0
\end{align}
The constrained optimization problem \eqref{opti2.min} and
\eqref{opti2.c} can be solved by standard optimization algorithms.

\section{Results of Testing simulations}
\label{sec:result}
We ran molecular dynamics (MD) simulations
on an Intel Xeon E5520 Processor with Gromacs
4.0.7 compiled by GCC 4.5. The testing systems used in our studies
contained 16,000 particles interacting via the standard Lennard-Jones
6-12 interaction.
The MD time step {was} $\Delta t = 0.002\,\tau$, where $\tau =
\sigma\sqrt{m/\varepsilon}$. NVT
simulations {were} performed by coupling the systems to the Nos\`e-Hoover
thermostat with a relaxation time $1\,\tau$. Short test runs of 10,000
steps, $r_1 = r_2-\sigma$ and $M=20$ {were} performed at different $r_2$ to provide an
estimate of the constants in the computational cost
expressions \eqref{tf} and \eqref{tn}. This process {lasted} for about an
hour. It is worthwhile to spend this time because it is short comparing
with the time costs of real simulations. Moreover, the constants can be
reused in all simulations on the same machine with the same density.
The error estimate \eqref{error.sr} ($m=6$ for Lennard-Jones interaction) {was} studied by the systems in the  gas
($T=1.20\varepsilon/k_B$, $\rho=0.05\sigma^{-3}$), liquid ($T=1.20\varepsilon/k_B$,
$\rho=0.80\sigma^{-3}$) and supercritical state ($T=1.34\varepsilon/k_B$,
$\rho=0.30\sigma^{-3}$). Target precisions 
$10^{-2}\varepsilon/\sigma$, $10^{-3}\varepsilon/\sigma$ and
$10^{-4}\varepsilon/\sigma$ {were} tested. To measure the real accuracies, the cutoff
radii equal to half the simulation boxes {were} employed and the resulting
forces served as the exact forces. The tuned parameters, accuracies
and computational costs
of the single- and twin-range cutoff methods
are listed in Tables \ref{tab1} and
\ref{tab2}.

\begin{table}[t]
  \centering
  \begin{tabular*}{0.85\textwidth}{@{\extracolsep{\fill}}cc|cccccc}\hline
    $\rho$ [$\sigma^{-3}$]&
    $\mathcal E^\ast$ [$\varepsilon/\sigma$]&
    $r_1$ [$\sigma$]&
    $d$  [$\sigma$]&
    $M$ &
    $\mathcal E_{real}$ [$\varepsilon/\sigma$]&
    $10^4\,T_{est}$ [\textsf s]&
    $10^4\,T_{real}$ [\textsf s]  \\\hline
         & $10^{-2}$ & 3.23 & 0.200 & 52 & $1.10\times 10^{-2}$ & 21.3  & 19.7 \\
    0.05 & $10^{-3}$ & 4.90 & 0.200 & 52 & $1.09\times 10^{-3}$ & 37.0  & 36.8 \\
         & $10^{-4}$ & 7.48 & 0.200 & 52 & $1.12\times 10^{-4}$ & 92.9  & 93.1 \\
    \hline
         & $10^{-2}$ & 3.91 & 0.180 & 44 & $1.17\times 10^{-2}$ & 85.1  & 88.0 \\
    0.30 & $10^{-3}$ & 5.87 & 0.126 & 31 & $1.43\times 10^{-3}$ & 266   & 268  \\
         & $10^{-4}$ & 8.91 & 0.100 & 25 & $1.86\times 10^{-4}$ & 914   & 851  \\
    \hline
         & $10^{-2}$ & 4.28 & 0.110 & 29 & $0.69\times 10^{-2}$ & 273   &251   \\
    0.80 & $10^{-3}$ & 6.45 & 0.081 & 21 & $0.64\times 10^{-3}$ & 900   &873   \\
         & $10^{-4}$ & 9.81 & 0.067 & 17 & $0.61\times 10^{-4}$ &3168   &3120  \\\hline
  \end{tabular*}
  \caption{
    The tuned parameters of the single-range cutoff method. System in the gas ($\rho=0.05\sigma^{-3}$, $T=1.20\varepsilon/k_B$), supercritical ($\rho=0.30\sigma^{-3}$, $T=1.34\varepsilon/k_B$) as well as liquid state ($\rho=0.80\sigma^{-3}$, $T=1.20\varepsilon/k_B$) were tested. $\mathcal E^\ast$ is the target accuracy while the $\mathcal E_{real}$ is the real error calculated from the simulation. $T_{est}$ is the computational expense estimated by \eqref{tsr}. $T_{real}$ is the real computational cost timed in the test simulations. The unit of the computational costs is second per step.
  }
  \label{tab1}
\end{table}

\begin{table}[t]
  \centering
  \begin{tabular*}{0.95\textwidth}{@{\extracolsep{\fill}}cc|ccccccc}\hline
    $\rho$ [$\sigma^{-3}$]&
    $\mathcal E^\ast$ [$\varepsilon/\sigma$]&
    $r_1$ [$\sigma$]&
    $r_2$ [$\sigma$]&
    $d$  [$\sigma$]&
    $M$ &    
    $\mathcal E_{real}$ [$\varepsilon/\sigma$]&
    $10^4\,T_{est}$ [\textsf s]&
    $10^4\,T_{real}$ [\textsf s]\\\hline
         &$10^{-2}$ & 2.83 & 3.69 & 0.200 & 52 & $1.08\times 10^{-2}$ & 19.7  &  19.3  \\
    0.05 &$10^{-3}$ & 4.24 & 5.61 & 0.200 & 52 & $1.03\times 10^{-3}$ & 30.9  &  33.0  \\
         &$10^{-4}$ & 6.05 & 8.24 & 0.126 & 33 & $1.02\times 10^{-4}$ & 69.7  &  77.4  \\
    \hline
         &$10^{-2}$ & 3.41 & 4.31 & 0.127 & 31 & $1.06\times 10^{-2}$ & 72.9  & 82.0   \\
    0.30 &$10^{-3}$ & 4.96 & 6.36 & 0.082 & 20 & $1.20\times 10^{-3}$ & 215   & 235    \\
         &$10^{-4}$ & 7.52 & 9.54 & 0.065 & 16 & $1.49\times 10^{-4}$ & 738   & 738    \\
    \hline
         &$10^{-2}$ & 3.76 & 4.64 & 0.080 & 21 & $0.85\times 10^{-2}$ & 238   & 228    \\
    0.80 &$10^{-3}$ & 5.60 & 6.90 & 0.057 & 14 & $0.76\times 10^{-3}$ & 767   & 793    \\
         &$10^{-4}$ & 8.59 &10.41 & 0.047 & 12 & $0.75\times 10^{-4}$ & 2742  & 2882   \\\hline
  \end{tabular*}
  \caption {
    The tuned parameters of the twin-range cutoff method. System in the gas ($\rho=0.05\sigma^{-3}$, $T=1.20\varepsilon/k_B$), supercritical ($\rho=0.30\sigma^{-3}$, $T=1.34\varepsilon/k_B$) as well as liquid state ($\rho=0.80\sigma^{-3}$, $T=1.20\varepsilon/k_B$) were tested. $\mathcal E^\ast$ is the target accuracy while the $\mathcal E_{real}$ is the real error calculated from the simulation. $T_{est}$ is the computational expense estimated by \eqref{tsr}. $T_{real}$ is the real computational cost timed in the test simulations. The unit of the computational costs is second per step. The last column gives the acceleration ratios of the real computational costs with respect to the real costs of the single-range cutoff method.
   }  
   \label{tab2}
\end{table}



In all cases presented, the constraint
\eqref{opti1.c1} is strictly satisfied, so the deviation of the real errors
from the target errors measures the quality of the error
estimates. In the gas state, the error estimates are sharp. In the
supercritical and liquid states, the error estimates do not deviate very
far from the real errors, and all of them fall in the range
$[\frac12\mathcal E_{real}, 2\mathcal E_{real}]$. In the supercritical
cases, the errors tend to be underestimated, while in the liquid
systems, the errors are somehow overestimated. The estimated RMS
errors are not exactly the real values, because some of
the assumptions in section \ref{sec:error} are not preserved in real
simulations.  The assumptions 3, 4 and 5
are obviously not satisfied because any two particles cannot overlap due to the repulsive
core of the Lennard-Jones interaction.
Moreover, they are also violated by the correlations between
particles due to the attractive dispersion term. 
It is possible to
include the pair distribution information (i.e. the radial
distribution function $g(r)$ in the homogeneous systems) to improve
the quality of the error estimates, but the estimates are then
\emph{posterior} rather than \emph{prior}, which is not convenient for
the parameter tuning.


From Table \ref{tab1} and \ref{tab2}, it is obvious that higher target
accuracies and larger system densities require more intensive parameters, i.e.
larger cutoff radii and higher neighbor list updating frequencies. In most gas
cases, the displacement $d$ hits the constraint
\eqref{opti1.c2}. In all cases shown, the $M$ are
larger than the usually used value ($M=10$, default in Gromacs). In the gas
state, the $M$ can be as large as 52.
The twin-range
cutoff method is always more efficient than the single-range cutoff method.
However, in
the liquid state, the twin-range cutoff method is a little bit less accurate
than the single-range cutoff method. So the benefit of using twin-range cutoff should be
discounted in the sense of the same accuracy.


To test the benefit we gain by the parameter tuning, we {used} a set of unoptimized
single-range parameters ($r=4.00\sigma$ and $M=10$) in the gas state
($\rho=0.05\sigma^{-3}$, $T=1.20\varepsilon/k_B$). The accuracy and efficiency
of these parameters are $\mathcal E=3.01\times 10^{-3}\varepsilon/\sigma$ and
$T=4.10\times10^{-3}\textsf s$, respectively. To compare the efficiencies, we
manually {adjusted} the target precisions $\mathcal E^\ast$ so that the real
accuracy of the tuned parameters is nearly the same as the corresponding
unoptimized parameters. The tuned single-range parameters are $r=4.09\sigma$ and
$M=52$ with computational cost $T=2.71\times10^{-3}\textsf s$. And the tuned
twin-range parameters are $r_1=3.54\sigma$, $r_2=4.68\sigma$ and $M=52$ with
$T=2.56\times10^{-3}\textsf s$. The tuned single- and twin-range
parameters save 33\% and 38\% computational costs, respectively. The same
unoptimized parameters result in $\mathcal E=6.10\times
10^{-3}\varepsilon/\sigma$ and $T=2.74\times10^{-2}\textsf s$ in the liquid
state ($\rho=0.80\sigma^{-3}$, $T=1.20\varepsilon/k_B$). At the same accuracy
level, the tuned single-range parameters are $r=4.36\sigma$ and $M=28$ with the
computational cost $T=2.76\times10^{-2}\textsf s$. And the tuned twin-range
parameters are $r_1=3.94\sigma$, $r_2=4.86\sigma$ and $M=19$ with
$T=2.76\times10^{-2}\textsf s$. In this case, the tuned parameters do not
improve the efficiency at all, because the unoptimized parameters are good
enough. Moreover, since the error estimate \eqref{error.sr} is not sharp in the
liquid state and the constants $c_1$, $d_1$, $c_2$ and $d_2$ in \eqref{tf} and
\eqref{tn} are not exactly measured, the tuned parameters are only nearly
optimal rather than really optimal. So the computational costs of the tuned
parameters can be trivially more expensive than the unoptimized ones. We also
{tested} a set of unoptimized twin-range parameters that are taken from one of our
former simulation settings: $r_1=2.5\sigma$, $r_2=4.0\sigma$ and $M=10$. In the gas
state, the error is $\mathcal E=4.48\times10^{-3}\varepsilon/\sigma$ with
$T=3.87\times10^{-3}\textsf s$. The tuned parameters are $r_1=3.28\sigma$,
$r_2=4.31\sigma$ and $M=52$ with the computational cost of
$T=2.29\times10^{-3}\textsf{s}$, 40\% cheaper. While in the liquid state, the
accuracy and computational cost of the unoptimized parameters are $\mathcal
E=2.91\times 10^{-2}\varepsilon/\sigma$ and $T=1.85\times 10^{-2}\textsf{s}$,
respectively. The tuned parameters are $r_1=3.12\sigma$, $r_2=3.81\sigma$ and
$M=27$ with the computational cost of $T=1.37\times 10^{-2}\textsf{s}$, 26\%
faster than the unoptimized parameters.



\section{Conclusions and open questions}
\label{sec:conclusion}

In this paper, an error estimate of the twin-range cutoff method was
developed.  The error estimate of the single-range cutoff method was
easily derived from the twin-range cutoff method, because the former
is only a special case of the later.  Equipped with both the error 
and the computational cost estimates, we proposed a work flow
that can automatically determine the nearly optimal parameters
demanding a certain target accuracy.  We verified the effectiveness of
the error estimate and parameter tuning algorithm by numerical
simulations of Lennard-Jones 6-12 system in gas, liquid and
supercritical state.  We also presented some examples to show, by
parameter tuning, to what extent the computational expense can be saved with
respect to the unoptimized parameters.

In the gas state, the error estimates are sharp. In the liquid
and supercritical state, the error estimates are reasonable: they
are always larger than half and smaller than twice of the real
errors.  The quality of the error estimates depends on the extent to which the
assumptions in section \ref{sec:error} are satisfied.  These
estimated force errors provide a quantitative description of the
undesirable cutoff effects introduced by the single- and
twin-range cutoff methods.
The parameter tuning algorithm enables an automatic searching of the
working parameters (i.e. the cutoff radii and the neighbor list updating
frequency) that reach a predetermined accuracy at nearly minimal
computational cost. Comparing with the unoptimized parameters,
the tuned parameters are always faster. The benefit ranged from
0\% to 40\%, depending on the quality of the unoptimized parameters.
We also demonstrated that the optimized twin-range cutoff method is
faster than the optimized single-range cutoff method, so the former is
recommended. Combining with the parameter tuning algorithm developed
for the long-range interactions \cite{wang2010optimizing}, all
non-bonded interactions can be calculated at a nearly optimal
efficiency.


The error estimate and the parameter tuning for inhomogeneous systems
were not considered by the present paper, 
because the
assumptions for the error estimate are worse violated. However, how to
handle the inhomogeneous systems is a very important open question in
this field, which requires further studies.

\section*{Acknowledgments}
The authors gratefully acknowledge Florian Dommert for our valuable
discussions and Jia Deng for his help editing our English.
P.Z. thanks the financial support by the National Natural Science
Foundation of China (50930003).



\end{document}